\documentclass[12pt]{article}
\newcommand{\D}{\Delta}
\renewcommand{\d}{\delta}

\renewcommand{\l}{\lambda}

\newcommand{\G}{\Gamma}
\newcommand{\g}{\gamma}
\newcommand{\e}{\epsilon}
\newcommand{\s}{\sigma}

\renewcommand{\i}{\iota}
\renewcommand{\a}{\alpha}

\newcommand{\m}{\mu}

\renewcommand{\r}{\rho}
\renewcommand{\s}{\sigma}
\renewcommand{\t}{\tau}
\newcommand{\p}{\pi}
\newcommand{\f}{\phi}

\renewcommand{\th}{\theta}
\renewcommand{\e}{\epsilon}
\newcommand{\ha}{\frac{1}{2}}
\renewcommand{\dim}{\textrm{dim}}

\newcommand{\be}{\begin{equation}}
\newcommand{\ee}{\end{equation}}
\newcommand{\bea}{\begin{eqnarray}}
\newcommand{\eea}{\end{eqnarray}}

\begin{document}

\begin{center}
\textbf{\Large Effective action and semiclassical limit of spin foam models}
\end{center}

\bigskip
\begin{center}
A. MIKOVI\'C\,\footnote{Member of the mathematical physics group at the University of Lisbon.} \\
Departamento de Matem\'atica,
Universidade Lus\'ofona de Humanidades e Tecnologias\\
Av. do Campo Grande, 376, 1749-024 Lisboa, Portugal\\
E-mail: amikovic@ulusofona.pt\\
\end{center}
\centerline{and}
\begin{center}
M. VOJINOVI\'C\\
Grupo de Fisica Matem\'atica da Universidade de Lisboa\\
Av. Prof. Gama Pinto, 2, 1649-003 Lisboa, Portugal\\
E-mail: vmarko@cii.fc.ul.pt
\end{center}

\centerline{}

\bigskip
\bigskip
\begin{quotation}
\noindent\small{We define an effective action for spin foam models of quantum gravity by adapting the background field method from quantum field theory. We show that the Regge action is the leading term in the semi-classical expansion of the spin foam effective action if the vertex amplitude has the large-spin asymptotics which is proportional to an exponential function of the vertex Regge action. In the case of the known three-dimensional and four-dimensional spin foam models this amounts to modifying the vertex amplitude such that the exponential asymptotics is obtained. In particular, we show that the ELPR/FK model vertex amplitude can be modified such that the new model is finite and has the Einstein-Hilbert action as its classical limit. We also calculate the first-order and some of the second-order quantum corrections in the semi-classical expansion of the effective action.}\end{quotation}

\bigskip
\bigskip
\noindent{\large\bf{1. Introduction}}

\bigskip
\noindent The problem of determining the classical limit of a spin foam model and the corresponding quantum corrections is one of the least understood and it represents the greatest obstacle for formulating a realistic spin foam model of quantum gravity. Certain clues have been obtained over the years, and the first one was the result obtained in \cite{ebcz} about the speed of convergence of the Euclidean Barret-Crane (BC) spin foam model state sum as a function of the edge amplitude. It was observed that a typical convergent BC state sum has the dominance of the small spin configurations. However, one can not obtain a sufficient information about the classical limit of a spin foam model from the partition function, and in \cite{rop} a study of the large-distance asymptotic behaviour of the graviton propagator for the BC model was started. It was discovered that the graviton propagator will have the correct large-distance asymptotics if the boundary spin-network wavefunction has a certain Gaussian form. However, in \cite{bcgp} it was pointed out that the tensorial structure of the BC model graviton propagator is not correct. This problem was caused by the absence of the intertwiners in the BC model. Without intertwiners, one cannot construct the complete Hilbert space of Loop Quantum Gravity (LQG) on a three-dimensional boundary. 

Furthermore, it was pointed out in \cite{mlqgp} that it is difficult to construct a physical spin-network wavefunction which has the Gaussian form used in \cite{rop} and it is not clear whether such a wavefunction exists. However, it was noticed in \cite{mlqgp} that if a physical wavefunction has the Rovelli Gaussian form in the limit of large spins, then such an asymptotics is sufficient to ensure the correct large-distance asymptotics for the graviton propagator. Further studies on the large-spin asymptotics of a class of physical wavefunctions which can be constructed as three-dimensional spin-foam state sums have revealed that their large-spin asymptotic form can be a Gaussian function, but it can not be of the Rovelli type, see \cite{mv1,mv3}. The source of this problem was identified in the large-spin asymptotics of the ELPR/FK spin-foam vertex amplitude \cite{mv3}. Namely, this asymptotics is of the form
\be W(j,\vec n) \approx \frac{N_+ e^{i\a S_{vR}(j,\vec n)} + N_- e^{-i\a S_{vR}(j,\vec n)}}{V(j)} \,,\label{va}\ee
where $j$ are the spins of the faces meeting in a given vertex, $\vec n$ are the corresponding coherent state vectors and $S_{vR}(j,\vec n)$ is the corresponding 4-simplex Regge action, see \cite{bare,barl}. $V(j)$ is a homogeneous function of order 12, $N_\pm$ are homogeneous functions of $j$ of order zero and $\a$ is a constant. On the other hand, it was shown in \cite{mv3} that a vertex amplitude whose asymptotics would be given by (\ref{va}) such that $N_- =0$ or $N_+ =0$ would give the correct graviton propagator asymptotics.

This result can be easily understood from the path-integral point of view, since the vertex asymptotics with $N_- =0$ gives the state sum which for large spins looks like the usual path-integral 
\be Z = \int {\cal D}\f \,e^{iS(\f)} \,,\label{epi}\ee
where $S(\f)$ is the Einstein-Hilbert action. On the other hand,
the asymptotics (\ref{va}) with $N_+ N_- \ne 0$ gives the state sum which for large spins looks like 
\be \tilde Z = \int {\cal D}\f \left[ e^{iS(\f)} +  e^{-iS(\f)}\right] \,,\label{cpi}\ee
and this is an unusual form of the path integral.

The authors of \cite{bs} argued that the path integral (\ref{cpi}) could still give the correct classical limit, if an appropriate boundary state is used. However, as argued in \cite{mv3}, it is difficult to prove that such a state exists. In this paper we will demonstrate that the usual path integral (\ref{epi}) gives the correct classical limit, and this will be done by using the effective action. At the heuristic level this can be seen from the background field method definition of the effective action
\be e^{i\G(\f)/\hbar} = \int {\cal D}\bar\f \,e^{iS(\f + \bar\f)/\hbar} \,,\label{bfma}\ee
see \cite{bfm} for details.
The definition (\ref{bfma}) implies that $S(\f)$ is the classical limit of $\G(\f)$ in the case of the usual path integral, while in the case of the path integral (\ref{cpi}) we obtain
$$ e^{i\G(\f)/\hbar} = \int {\cal D}\bar\f \left[ e^{iS(\f + \bar\f)/\hbar}+ e^{-iS(\f + \bar\f)/\hbar}\right]\,.$$
It is clear that this expression will not give $S(\f)$ as the classical limit of $\G (\f)$.
However, in order to have the state sum which for large spins looks like (\ref{epi}), we need to change the vertex amplitude such that the large-spin asymptotics is given by 
\be \tilde W(j,\vec n) \approx \frac{ e^{i\a S_{vR}(j)}}{ V(j)} \,.\label{nva}\ee

In this paper we will show that this can be done in the case of the EPRL/FK class of spin foam models, and that the length-Regge action can be obtained in the large-spin limit. In section 2 we discuss the known approaches to the problem of the classical limit of spin foam models and conclude that the effective action approach based on the background field method is the most feasible. We give the definition of the effective action for spin foam models and in section 3 we study the case of the Ponzano-Regge model. We show that the corresponding effective action has the desired classical limit if the vertex amplitude is redefined such that the large-spin asymptotics is proportional to the exponential function of the vertex Regge action times the imaginary unit. We calculate the corresponding effective action in the large-spin limit and show that the Regge action is the dominant term in the semi-classical expansion. We also formulate a finite Lorentzian Ponzano-Regge model such that its classical limit is the Einstein-Hilbert action. In section 4 we study the case of the Lorentzian ELPR/FK spin foam model and show that the vertex amplitude can be redefined such that its large-spin asymptotics is proportional to the exponential function of a constant times the vertex Regge action. The corresponding effective action can be calculated in the large-spin limit by using the same techniques as in the three-dimensional case, and the result is the length-Regge action plus the quantum corrections. In section 5 we present our conclusions.

\bigskip
\bigskip
\noindent{\large\bf{2. Classical limit of spin foam models}}

\bigskip
\noindent Spin foam models of quantum gravity in $D=3$ and $D=4$ spacetime dimensions are described by a partition function of the form
\be Z = \sum_{j,\i} \prod_f W_2 (j_f) \prod_l W_1 (\i_l) \prod_v W_0 (j_{f(v)},\i_{l(v)})\,,\label{gsfz}\ee
where $j=(j_1,\cdots,j_f,\cdots,j_F)$ and $\i = (\i_1,\cdots,\i_l,\cdots,\i_L)$ are labels of a spin foam whose two-complex $\s$ is dual to the simplicial complex obtained by triangulating the spacetime manifold. The faces $f$ and the links $l$ of $\s$ carry the labels $j_f$ and $\i_l$, respectively, and these labels are irreducible representations and the corresponding intertwiners of the group $Spin(D)$. Since $Z$ is a complex number, it is not possible to extract the classical limit from it and one needs to analyze the boundary wavefunctions or to analyze the effective action.

A boundary wavefunction $\Psi (s)$ is associated to a boundary spin network $s=(\g,j_b,\iota_b)$, where $\g$ is the boundary one-complex of $\s$ and $(j,\i)$ are the corresponding labels. $\Psi(s)$ is constructed from (\ref{gsfz}) such that the summation is restricted to spin foams whose boundary spin network is $s$. Therefore
\be \Psi (s) = \sum_{j,\i} \prod_f \tilde W_2 (j_f) \prod_l \tilde W_1 (\i_l) \prod_v W_0 (j_{f(v)},\i_{l(v)})\,,\label{hhw}\ee
where the amplitudes $\tilde W$ are the same as the amplitudes $W$ for the faces and the links not belonging to the boundary, while for the boundary faces and links there is a choice which ensures good gluing properties, see \cite{brr}. 

Note that the construction (\ref{hhw}) gives just one boundary state 
\be |\Psi\rangle = \sum_s \Psi (s) |s\rangle \,,\label{bs}\ee
while we know from the canonical LQG that there are many different physical states. Especially important physical states are those which describe flat or constant curvature
spatial manifolds. Therefore the definition (\ref{hhw}) has to be changed such that the information about
the background triads $E_0(x)$ is included, where $x$ is a spatial coordinate. In the case of Euclidean canonical LQG one can show that such a wavefunction has a form similar to (\ref{hhw}), but the spin network $s$ is replaced by a spin network $\tilde s$ where $\tilde s$ is $s$ with edge insertions $\mu_l (E_0(l))$, where $E_0(l) = \int_l E_0 (x) dx$, see \cite{mikfw}. The insertion functions can be chosen freely, and an appropriate choice are the Gaussians centered around $E_0 (l)$. In the case of a flat geometry, all $E_0(l)$ can be taken to be approximately the same, and the corresponding area of a triangle is proportional to $j_0$. This parameter will set the length scale, so that one introduces the insertions into the boundary spin network of (\ref{hhw}) which will be functions of $j_0$.

Given such a $\Psi(s,j_0)$, there is the corresponding connection wavefunction $\Psi_0 (A)$, which can be obtained by the loop transform. By writing 
$$\Psi_0 (A) = R(A)\,e^{iS(A)/\hbar} \,,$$ 
one would have to show that
$$S(A)= S_0(A) + \hbar S_1(A) + O(\hbar^2)$$ 
where $S_0(A)$ satisfies the Hamilton-Jacobi equation for canonical general relativity (GR). It is obvious that this is an extremely difficult way to obtain the classical limit.

An easier approach would be to calculate the graviton correlation functions for the boundary state (\ref{bs}) with insertions, so that 
$$G_n (x_1,\cdots,x_n)=\sum_{s,s'} \Psi_0^*(s) \Psi_0(s')\langle s| \hat h(x_1) \cdots \hat h(x_n)|s'\rangle \,,$$ 
where $\hat h$ is the graviton operator. This was the approach started by Rovelli \cite{rop}, and it can be shown that $G_2$ has the correct large-distance asymptotics if 
\be \Psi_0 (s) \approx N \exp \left( -{1\over j_0}\sum_{l,l' \subset\g} \a_{ll'}(j_l - j_{0})(j_{l'} - j_{0})  \right)\,,\label{ga}\ee
for large spins, where $\a$ is a constant matrix \cite{mlqgp}.

However, the correct asymptotics of $G_2$ does not guarantee that the classical limit of a spin foam model is general relativity. Namely, one has to show that all correlation functions $G_n$ correspond to the ones for the EH action in the classical limit. This is equivalent to demonstrating that the effective action
\be \G[h] = \sum_{n\ge 2} c_{n} \int dx_1 dt_1 \cdots \int dx_n dt_n \tilde{G}_n (x_1,t_1,\cdots, x_n,t_n)h(x_1,t_1) \cdots h(x_n,t_n) \,,\label{hea}\ee
has the classical limit which is given by the Einstein-Hilbert action,
where $(x,t)$ is a spacetime point coordinate and $\tilde G_n$ is the extension of $G_n$ when $t_k \ne  t_l$. The correlation function approach is less difficult then the wavefunction approach, but it still requires a lot of work. 

A simpler method to compute the effective action would be a method where $\G$ is given as a functional of the spacetime metric $g$ rather then the functional of $h = g -\eta$, where $\eta$ is the flat metric. The background field method (BFM) of computing the effective action in quantum field theory (QFT) \cite{bfm}, is a method convenient for this purpose. In the case of quantum gravity, the BFM approach suggests the following relation
\be e^{i\G(g)/\hbar}= \int {\cal D}h\, e^{iS(g+h)/\hbar} \,,\label{qgea}\ee
where $S(g)$ is the Einstein-Hilbert action. The expression (\ref{qgea}) is formal and has to be defined, and in the perturbative QFT approach it amounts to gauge fixing of the diffeomorphism gauge symmetry and implementing a regularization/renormalization procedure. This has to be done because the theory allows arbitrary short distances and hence the infinities appear. Since GR is a non-renormalizible theory, the corresponding effective action can not be determined uniquely.

In the case of spin foam models, the problems with QFT infinities are avoided because the theory has a natural short distance cut-off. Namely, the basic degrees of freedom are the $SU(2)$ spins $j_f \in {\bf N}/2$, and these are essentially the areas of the corresponding triangles, since $A_f \propto l_P^2 \sqrt{j_f (j_f +1/2)}$. Hence there is a short-distance cut-off of order of the Planck length $l_P$. There is no need for a gauge-fixing procedure, since $j_f$ are triangle areas, and these are diffeomorphism invariant. The only infinites which can appear are in the large $j_f$ region, which correspond to large-distance infinities (also known as the infra-red infinities in QFT), but these can be easilly dealt with, by introducing the appropriate negative powers of $j_f$ in the spin-foam amplitude, see \cite{mv4}.
 
The path integral (\ref{qgea}) takes the following form in the case of spin foam models
\be e^{i\G(j,\i)}= \sum_{j',\i'}\, \prod_f W_2(j_f+j'_f) \prod_l W_1 (\i_l + \i'_l )\prod_v A (j_{f(v)}+ j'_{f(v)},\i_{l(v)} +\i'_{l(v)}) \,,\label{sfea}\ee
where the spin-foam two complex is closed, $(j,\i)$ is a configuration representing the background or the classical values of the spin-foam labels, while the summation is over the fluctuations $(j',\i')$ around the classical background $(j,\i)$. The calculation of $\G$ will simplify if we take that the background spins are large and that $(j,\i)$ is a stationary point of $S(j,\i)$, where
\be e^{S(j,\i)} = \prod_f W_2(j_f) \prod_l W_1 (\i_l  )\prod_v A (j_{f(v)},\i_{l(v)}) \,\label{sfamp}\ee
is the partition function spin-foam amplitude. The latter condition is also used in the QFT version of the BFM approach, where it takes the form of background metric being a solution of the Einstein equations.

As discussed in the introduction, the vertex amplitude $A$ should be a function of the $W_0$ amplitude such that $A$ has the asymptotics 
\be A (j,\i ) \approx \frac{e^{i\a S_{vR}(j)}}{V_p (j)}\,\label{expa}\ee
for $j\to\infty$, where 
\be S_{vR} = \sum_{f\supset v} j_f \th_{f} \ee
is the vertex Regge action and $\th_f$ are the dihedral angles, while
$V_p (j)$ should be a homogeneous function of order $p >0$. The role of the function $V_p (j)$ is to ensure that the state sum (\ref{sfea}) is convergent, see \cite{mv4}.

The requirement (\ref{expa}) is essential, since it will give the Regge action
\be S_R = \sum_f j_f \d_f \label{ra}\ee
as the classical limit of the effective action. Namely, if we take the background spins to be large, then we can use the asymptotic formula (\ref{expa}) in the state sum (\ref{sfea}), which then produces the Regge action in the exponent due to the identity
\be S_R = \sum_v S_{vR} + 2\pi \sum_f  k_f j_f \,,\label{ravr}\ee
where $k_f$ are integers.

It will be important to notice that the deficit angle $\d_f$ is given by
\be \d_f = 2\p - \sum_{v\subset f} \th'_{fv}\,,\label{sdiha} \ee
for a spacelike face $f$, where $\th'_{fv} = \pi -\th_{fv}$ is the interior dihedral angle for the 4-simplex $\s_v$ dual to $v$. A spacelike $f$ means that the dual triangle $\D_f$ is timelike and belongs to $\s_v$. When $f$ is timelike, which means that $\D_f$ is spacelike, then
\be \d_f = \sum_{v\subset f} \Theta_{fv}\,,\label{tdih} \ee
where $\th_{fv} =\Theta_{fv}$ and $\Theta_{fv}$ is the boost parameter between the normal vectors of the two tetrahedrons of $\s_v$ which share the triangle $\D_f$, see \cite{brfx}.

\bigskip
\noindent{\large\bf{4. The three-dimensional case}}

\bigskip
\noindent Let us explore first the effective action given by (\ref{sfea}) in the simpler case of three-di\-men\-si\-o\-nal (3d) spin foam models. Consider the Ponzano-Regge model partition function
\be Z_{PR} = \sum_j \prod_f (-1)^{2j_f}\dim j_f \prod_v W (j_{f(v)})\,(-1)^{j_1(v) + \cdots + j_6(v)} \,,\label{prs}\ee
where $W$ is the $6j$-symbol, see \cite{prm}. The immediate problem with $Z_{PR}$ is that it is divergent, so that it has to be regularized. This can be done by introducing a maximum spin, or by dividing $W$ by an appropriate power of the product of the dimensions of the vertex spins. The later regularization will be more convenient for our purposes. Either way one loses the triangulation independence, but we will see that this is not going to be a problem for our purposes. 

The next problem is that
$$ W (j) \approx {\cos\left[S_{vR}(j)\right] \over \sqrt{V(j)}} \,,$$
for large spins, where $V(j)$ is the volume of the vertex tetrahedron. According to our approach we then need to change the vertex amplitude such that the new asymptotics is given by (\ref{expa}). In order to achieve this consider
\be \tilde{W} = \sqrt{V} W + \sqrt{VW^2 -1} \,.\label{6jr}\ee
It is easy to show that $\tilde W \approx e^{iS_{vR}}$ for large spins since (\ref{6jr}) implies
$$ W = {\tilde W + (\tilde W)^{-1}\over 2\sqrt{V}}\,.$$ 

Let us introduce a modified vertex amplitude
\be A(j) = \frac{\tilde W (j)}{\sqrt{V}\prod_{k=1}^6 (\dim j_k)^{p'}} \,.\label{mva}\ee
Then $A(j)$ will have the asymptotic form (\ref{expa}) with $p = 6p' + \frac{3}{2}$, when all of the six spins $j$ are large. The parameter $p'$ has to be chosen such that the state sum (\ref{sfea}) is finite.
This can be done because $\tilde W / \sqrt V$ is a limited function. Then exists $M>0$ such that
\be |A(j)| < \frac{M}{\prod_k (\dim j_k )^{p'}} \,.\ee
Consequently
\be |Z_p |< N \sum_j \prod_f (\dim j_f)^{1-p'n_f} \le N \sum_j \prod_f (\dim j_f)^{1- 2p'} \,.\label{3df}\ee
where $Z_p$ is the partition function associated to the state sum (\ref{sfea}). The $n_f$ denotes the number of vertices of a face $f$ and since $n_f \ge 2$, we obtain the last inequality. The last sum in (\ref{3df}) will be convergent for $p' > 1$. Therefore $Z_p$ will be convergent for $p' >1$. One can find a better estimate for the lower bound for $p'$ by using a better estimate for $\tilde W / \sqrt V$, but for us it is important that such $p'$ exist and that their values are independent from the triangulation.

We are interested in calculating the effective action when all the background spins in the state sum (\ref{sfea}) become large. Then one can approximate each vertex amplitude in (\ref{sfea}) by using (\ref{expa}), since all the spin labels $j+j'$ are large. Consequently
\be e^{i\G(j)} \approx \,N' \,  \sum_{j'}  \prod_f  (j_f +j_f')^{1 - p_f n_f}\, e^{iS_R (j+j')} \,,\label{lsea}\ee
where we used $A_2 (j+j') \approx 2 (j+j')$ and $p_f = p' + 3/2$. Note that the sign factors in the face and the vertex amplitudes, see (\ref{prs}), combine with the sum of the vertex Regge actions such that the Regge action is obtained in the exponent of (\ref{lsea}).

The main contribution in the state sum (\ref{lsea}) comes from $j'_f << j_f$, since the weights $(j_f +j_f')^{1 - p_f n_f}$ are maximal for $j_f' =0$ and drop-off as negative powers of $j'_f$. We can then use
$$ (j+j')^{-m} = j^{-m}\left(1- \frac{j'}{j} \right)^{-m} = j^{-m}\left[ 1 -m {j'\over j} +m(m+1) {j'^2 \over 2 j^2} + \cdots\right]\,, $$
which is valid for $j'< j$. Consequently
\be e^{i\G(j)} \approx N \,  \sum_{j'} e^{ iS_R (j+j')} \prod_f j_f^{-m_f}\left[ 1 -m_f 
{j'_f \over j_f}  + m_f (m_f +1) {{j'}_f^2 \over j_f^2 } + \cdots\right]\,,\ee
where $m_f = n_f p_f - 1$ is a positive number.

Let us choose the background spins $j_f$ such that they correspond to a stationary point of the Regge action $S_R (j)$. This is a standard procedure in the case of QFTs, and the idea is to simplify the calculation, since the stationary point restriction of the background spins does not affect the important features of the effective action. Therefore we can use the approximation
$$ S_R (j+j') \approx S_R (j) + \frac{1}{2}\sum_{f,f'} S''_{R\,ff'}(j)j'_f j'_{f'} \,,$$
where $S''_{R\,ff'}(j)$ is the Hessian matrix for $S_R (j)$. Consequently
\be e^{i\G(j)} \approx N \,   e^{ iS_R (j)-\sum_f m_f \ln j_f}\sum_{j'}e^{i\langle S''_R (j)j'j'\rangle/2} \prod_f \left( 1 -m_f {j'_f\over j_f} + \cdots\right)\,,\label{spa}\ee
where $\langle S''_R (j) j' j'\rangle =\sum_{f,f'} S''_{R\,ff'}(j)j'_f j'_{f'}$.
The sum in (\ref{spa}) can be approximated by an integral over $x_f = j'_f/j_f$ variables, and this integral will be given as a sum of the integrals of the following type
$$ \int d^F x \, x_1^{n_1} \cdots x_F^{n_F} \exp\left( \frac{i}{2}\sum_{m,n}S''_{R\,mn}x_m x_n\right) \,.$$
These integrals can be calculated by taking the derivatives of  the generating function
$$I(\mu) =  \int d^F x \, \exp\left( \frac{i}{2}\sum_{m,n}S''_{R\,mn}x_m x_n + \sum_m \m_m x_m\right) \,,$$
at $\m =0$, where $I(\m)$ is given by
\be I(\m) = (2\p i)^{F/2} \,{\exp\left(i \m^T (S''_R)^{-1} \m/2\right)\over\sqrt{\det(S''_R)}}\,, \label{genf}\ee
and $\m^T = (\m_1,\cdots,\m_F )$.

This calculation can be simplified by using
$$ (1+x)^{-m} = e^{-m\log(1+x)} = e^{-mx + mx^2/2 + \cdots}$$
so that 
\be e^{i\G(j,\iota)} \approx \,N \,e^{ -\sum_f m_f \log j_f  + iS_R (j)}\,\sum_{j'}e^{-\sum_f m_f j'_f/j_f + \frac{i}{2} \sum_{f,f'} \tilde S''_{R\,ff'} j'_f j'_{f'} } \,,\ee
where 
$$\tilde S''_{R\,ff'} =  S''_{R\,ff'} - i\d_{f,f'}{m_f \over j_f^2} \,.$$ 

Then by using (\ref{genf})
\be e^{i\G}
\approx  N' \, \exp \left( - \sum_f m_f\log j_f + iS_R (j)\right) 
 {\exp\left(i\sum_{f,f'}m_f m_{f'}{\tilde G_{ff'}(j)\over 2j_f j_{f'}}\right)\over (\det \tilde S''_R (j))^{1/2} }\ee
where $\tilde G$ is the inverse matrix of $\tilde S''$. Consequently
\be e^{i\G}\approx N'\exp \left( - \sum_f m_f\log j_f + iS_R (j)-\frac{1}{2}Tr\log \tilde S''_R (j)+ i\sum_{f,f'}m_f m_{f'}{\tilde G_{ff'}(j)\over 2j_f j_{f'}} \right)\,.\label{efap}\ee

The equation (\ref{efap}) implies that
\bea \G(j) &\approx&  S_R (j) + i\sum_f m_f \log j_f +\frac{i}{2}Tr\log \tilde S''_R (j) -i \log N' + \sum_{f,f'}m_f m_{f'}{\tilde G_{ff'}(j)\over 2j_f j_{f'}}\cr 
&\approx& S_R (j)  + i\sum_f m_f \log j_f +\frac{i}{2}Tr\log S''_R (j) -i \log N' + O(1/j)\,,\label{efac}\eea
where $O(1/j)$ denotes the terms which scale as $1/j$ when $j_f \to j j_f$. When necessary, it will be understood that notation $O(j^{m})$ also includes the subleading terms.

Note that
$S_R = O(j)$, while $\log j_f$ and $Tr\log S''_R$ are of $O(\log j )$. Therefore (\ref{efac}) implies
$$\G(\l j ) \approx \l\, \G_0(j) + (\log\l)\, \G_1 (j) + \G_2(j) + \l^{-1}\, \G_3 (j) + O(1/\l^2)\,, $$
in the limit $\l\to\infty$. The dominant term in the large-spin limit will be $\G_0$, which is the Regge action (\ref{ra}). Hence we can say that the classical limit of the effective action is the Regge action, which is a discretization of the Einstein-Hilbert action. This means that if we start refining the spacetime simplicial complex, $\G_0$ will tend to the EH action.

Note that the quantum correction terms in (\ref{efac}) are imaginary numbers, while the effective action should be a real function. The same problem appears in QFT, since in that case
$$ e^{i\G (\f)/\hbar} = \int {\cal D} \varphi \,e^{iS(\f + \varphi)/\hbar} \,.$$
The stationary phase approximation then implies
$$ \G(\f) \approx  S(\f) + i\,\frac{\hbar}{2}\, Tr\log S''(\f) + O(\hbar^2) $$
for a stationary point $S'(\f) =0$, so that $\G(\f)$ is not a real action. This problem is resolved by resorting to the Wick rotation. Namely, by performing the Wick rotation $t\to i t$, where $t$ is the time coordinate, one passes to the theory in the Euclidean metric and defines the effective action which is real
$$ e^{-\G (\f)/\hbar} \approx \int {\cal D} \varphi \,e^{-S(\f + \varphi)/\hbar} \,.$$
Consequently
\be \G(\f) \approx S(\f) + \frac{\hbar}{2}\,Tr\log S''(\f) + O(\hbar^2) \,.\label{eucea}\ee
Then in the Euclidean effective action (\ref{eucea}) one replaces the Euclidean metric with the Minkowski metric, and obtains a real effective action. In the case of  spin foam models, there are no spacetime coordinates and there is no background metric, so that one cannot perform the Wick rotation. However, note  that the procedure used in QFT is a redefinition of a complex function $\G = \G_0 + i\G_1$ into a real function $\G_0 + \G_1$. Therefore we can define a real effective action by
\be \G \to Re\,\G + Im\,\G \,.\label{rea}\ee
The definition (\ref{rea}) is not unique, since one can also use $Re\,\G - Im\,\G$. This ambiguity can be only resolved by an experiment, but we will use (\ref{rea}), because it agrees with the QFT theory sign. Therefore for large spins we obtain
\be \G(j) \approx  S_R (j) + \sum_f m_f \log j_f + \frac{1}{2}Tr\log S''_R (j) + O(1/j) \,.\label{3dea}\ee

We would like to make the following remarks. The partition function $Z_p$, which corresponds to the modified vertex amplitude (\ref{mva}) is finite,  but it is not a topological invariant. This is not a problem, since our goal is not constructing manifold invariants, but obtaining a quantum theory of gravity whose classical limit is general relativity. This is achieved by requiring that the state sum (\ref{sfea}) is finite and that the classical limit of the effective action is the Regge action (\ref{ra}). Although our construction is triangulation dependent, it still leads to a topological theory in the continuum limit. Namely, if we start refining the triangulation, the Regge action will become the EH action, which defines a topological theory in three spacetime dimensions. 

The trace-log term in (\ref{3dea}) is a discretization of the usual trace-log term from QFT. Since the QFT trace-log term is divergent, the spin foam version can be considered as a regularization 
of the QFT counterpart. In contrast, the $m \log j$ terms in (\ref{3dea}) do not have an analog in the QFT case, and their presence is a feature of the model. It is not clear what is the smooth limit of the $m\log j$ terms and whether their presence is a good or a bad feature of the model, since we do not know experimentally what are the quantum gravity corrections.

Note that one can define a spin foam model with a simpler vertex amplitude than (\ref{mva})
\be \tilde A (j) = \frac{e^{iS_{vR} (j)}}{\prod_k (\dim j_k)^p} \,.\label{3dsa}\ee
The corresponding state sum would look like the path integral for a Regge gravity model where the lengths of the edges of a triangulation are positive half-integers. One can enforce the triangle inequalities by inserting the dual-edge (triangle) amplitudes proportional to the theta spin network evaluation, so that
\be \tilde Z = \sum_{j} \prod_f (-1)^{2j_f}\dim j_f \prod_l \theta(j_{f(l)}) \prod_v \tilde A (j_{f(v)})(-1)^{j_1(v) +\cdots + j_6 (v)} \,.\label{nem}\ee
The corresponding effective action will be also given by the expression (\ref{3dea}).

The spin foam model defined by the $\tilde A$  amplitude can be easily extended to the Lorentzian case, simply by replacing the vertex Regge action $S_{vR}$ in (\ref{3dsa}) by its Lorentzian analog. We will label the edges of the triangulation with the unitary irreps $j_f$ from the discrete series of representations of $Spin(1,2)= SL(2,R)$, which makes sense if the edges are spacelike. Then the deficit angle is given by the 3d analog of (\ref{tdih}), so that the sum of the vertex Regge actions will be equal to the Regge action. Therefore we will not need the sign factors in the face and vertex amplitudes, which were necessary in the Euclidean case, in order to obtain the Regge action in the classical limit of the effective action. Hence one can define a finite Lorentzian 3d quantum gravity spin foam model whose partition function is given by
\be \tilde Z_L = \sum_{j} \prod_f (2j_f +1) \prod_l \theta(j_{f(l)}) \prod_v \tilde A (j_{f(v)}) \,.\label{l3dqg}\ee
The corresponding effective action will given by (\ref{3dea}) in the large-spin limit so that in the smooth spacetime limit one will obtain the EH action.

\bigskip
\bigskip
\noindent{\large\bf{5. The four-dimensional case}}

\bigskip
\noindent We will consider the ELPR/FK spin foam models \cite{elpr,fk}, since this is the only class of four-dimensional (4d) spin foam models that has a well-defined LQG theory on a 3d boundary. The partition function is given by (\ref{gsfz}) such that
$W_2 (j) = \dim\, j$, $W_1 (j,\i) = 1$ and 
$$ W_0 (j,\i) = \sum_{n_1 \ge 0,\cdots,n_5 \ge 0}\prod_{a=1}^5\int_0^{+\infty} d\r_a (n_a^2 + \r_a^2 ) f^{\i_a}_{n_a \r_a}(j) \,W_{15}(2j_{bc},2\g j_{bc}; n_b,\r_b ) \,,$$
where $\g$ is the Barbero-Immirzi parameter, $W_{15}$ is the $15j$-symbol for the unitary representations $(n,\rho)$ of the Lorentz group and $f$ are the fusion coefficients, see \cite{elpr} for the details. $W_2 (j)$ was originally chosen to be a quadratic function \cite{elpr}, but it has been recently argued in \cite{brr} that the linear weight is more appropriate. In any case, the essential features of the effective action are the same.

A more convenient form of $Z$ is
$$ Z = \sum_{j,\vec n} \,\,\prod_f \dim\,j_f \, \prod_v  W (j_{f(v)},\vec n_{lf(v)}) \,,$$
where each $\i_l$ in a spin foam from the sum (\ref{gsfz}) is replaced by four unit 3-vectors $\vec n_{lf}$. An $\vec n_{lf}$ is a 3-vector orthogonal to the triangle dual to face $f$, such that this triangle belongs to the tetrahedron dual to a link $l$.

For the purposes of calculating the effective action we only need to now the asymptotics of $W(j,\vec n)$ when all the spins $j$ are large. This asymptotics is given by
\be W(j,\vec n) \approx {N_+(\a) e^{i\a S_{vR}(j,\vec n)} + N_-(\a) e^{-i\a S_{vR}(j,\vec n)} \over V(j)} \label{eprlva}\ee
where 
$\a =\g$ when the 4-simplex boundary geometry is Lorentzian, while $\a =1$
when the 4-simplex boundary geometry is Euclidean, see \cite{barl}. The real numbers $N_\pm(\a)$ are $O(1)$ functions of the spins, while $V = O(j^{12})$. There are also degenerate configurations of spins for which $W(j,\vec n) \approx N(j)/V(j)$ where $N(j) = O(1)$ and $V(j) = O(j^{12})$, but their contribution to the state sum is negligible. Otherwise, $W$ drops off faster than any power of $1/j$.

In order to obtain a correct $\G_0$ we need to redefine the vertex amplitude. Let us introduce a new vertex amplitude $\tilde W (j,\vec n )$ such that
\be \tilde W  = {V W + \sqrt{(V W)^2 -4N_+ N_-}\over 2N_+} \,.\ee
This formula follows from the following relation between $W$ and $\tilde W$
$$W = {N_+ \tilde W + N_- (\tilde W )^{-1} \over V(j)}\,,$$
which ensures that
$$ \tilde W (j,\vec n) \approx e^{i\g S_{vR}(j,\vec n)} \,,$$
for large spins. One can then define
\be A(j,\vec n) = { \tilde W(j,\vec n)\over V(j) \prod_{f} (\dim j_f)^p } \,,\label{gva}\ee
where $p$ is sufficiently large such that $Z_p$ is convergent. Such a $p$, which does not depend on the triangulation, can be always arranged, as shown in \cite{mv3}.

The amplitude (\ref{gva}) has the desired asymptotics and it gives a finite partition function. We can now use the formula (\ref{sfea}) for the effective action, which takes the form
\be e^{i\G(j,\vec n)}= \sum_{j',\vec n'}\, \prod_f [2(j_f+j'_f)+1] \prod_v A (j_{f(v)}+ j'_{f(v)},\vec n_{fl(v)} + \vec n'_{fl(v)}) \,.\label{fdea}\ee
The labeling of the ELPR/FK model is consistent with a spacelike triangulation. In that case $k_f =0$ in the formula (\ref{ravr}),  so that when $(j_1,\cdots,j_F)\to (\infty,\cdots,\infty)$ we obtain
\bea e^{i\G(j,\vec n)}&\approx& N \sum_{j',\vec n'}\, \prod_f (j_f+j'_f) \prod_v {{e^{-i\g S_{vR} (j_{f(v)}+ j'_{f(v)},\vec n_{fl(v)} + \vec n'_{fl(v)})}}\over V(j)\prod_f ( j_f + j_f')^p}\cr
&\approx& N\sum_{j',\vec n'}\, \prod_f (j_f+j'_f)^{1-p_f m_f} e^{i S_R (j+ j',\vec n + \vec n')}\,,\label{rap}\eea
where $p_f\ge p$ and $m_f$ is the number of vertices belonging to a face $f$. Note that the contribution to $e^{i\G}$ of the configurations $(j+j',n+n')$ which are not geometric is negligible compared to (\ref{rap}), since $A(j+j',n+n')$ decreases exponentially with large spins. The contribution of degenerate geometric configurations is also negligible, since one sums over a lower-dimensional sub-space in the space of spins.

Let $j$ and $\vec n$ be the background spin foam labels such that $(j,\vec n)$ is a stationary point of $S_R (j,\vec n)$ and all the 4-complexes have the Lorentzian geometry. Then we can use the formulas from the 3d case, and we obtain
\bea e^{i\G(j,\vec n)}&\approx& N \sum_{j',\vec n'} e^{ -\sum_f c_f \log(j_f + j_{f'}) + i S_R (j,\vec n) + \frac{1}{2}\langle S''_{R\,jj}j'j'+ 2S''_{R\,j n}j'\vec n' + S''_{R\, n n}\vec n'\vec n'\rangle}\cr
&\approx& N \sum_{j',\vec n'} e^{ -\sum_f c_f [\log(j_f) + \frac{j_{f'}}{j_f} ] + i S_R (j,\vec n) + \frac{1}{2}\langle \tilde S''_{R\,jj}j'j'+ 2S''_{R\,jn}j'\vec n' + S''_{R\,nn}\vec n'\vec n'\rangle}\,,\label{stpexp}\eea
where $c_f = p_f m_f -1$ and
$$\tilde S''_{R\,ff'} = S''_{R\,ff'} -i \frac{c_f }{ j_f^2}\d_{ff'} \,.$$

By performing the Gaussian integrations in (\ref{stpexp}) we obtain
\be e^{i\G(j,\vec n)}\approx N'\,e^{ -\sum_f c_f \log(j_f) + i S_R (j,\vec n)-\frac{1}{2}\,Tr \log \tilde S''_R (j,\vec n) + \ha\sum_{f,f'} \tilde c_f  \tilde c_{f'} G_{R\,ff'}(j,\vec n)} \,,  \label{4dstpa}\ee
where 
$$ \tilde S''_R = \left(\matrix{\tilde S''_{R\,jj} \,\, S''_{R\,jn} \cr S''_{R\,jn} \,\, S''_{R\,nn}}\right)\,, $$
$G_{R\,ff'}$ is an element of the $jj$ block of the matrix $(S''_R)^{-1}$ and $\tilde c_f = c_f /j_f$. 

The equation (\ref{4dstpa}) implies
\be \G(j,\vec n) \approx  S_R (j,\vec n) +\sum_f c_f \log \,j_f + \frac{1}{2}\,Tr \log \tilde S''_R (j,\vec n) + \sum_{f,f'} c_f  c_{f'} \frac{G_{R\,ff'}(j,\vec n)}{2j_f j_{f'}} \,,\label{scea}\ee
where we have used $\G\to Re\,\G + Im\,\G$ and we have omitted the constant $\log N'$. As in the 3d case, the dominant term is $S_R$, which is of $O(j)$, while the other terms are of subleading orders, namely of $O(\log j)$ and $O(1/j)$, respectively. 

The equation (\ref{scea}) implies that the classical limit of the ELPR/FK effective action is the area-Regge action
\be  S_{R}(j,\vec n) =\g\sum_f j_f \d (j_f , \vec n)\,.\label{ar}\ee
However, we can require that the background spin foam $(j,n)$ is also a stationary point of the partition function amplitude, see (\ref{sfamp}). This requirement imposes further restrictions on the background spin foam, and it was shown in \cite{cf,mp} that such stationary points correspond to Regge geometries when $j_f$ are all large. This means that there is an assigment of the edge lengths $L_\e$, satisfying the triangle inequalities, such that $j_f \propto A_f (L)$ and $\vec n = \vec n(L)$, where $A_f$ is the area of the triangle dual to a face $f$. 

More precisely, the stationary point equations imply that the four normals $\vec n$ associated to a tetrahedron $\t$, which are determined by the four-geometry of a four-simplex $\s$ which contains $\t$, are the same as the four normals $\vec n'$ determined by the four-geometry of another $\s'$ containing $\t$. Consequently, the length of an edge $\e$ of $\t$ is the same when calculated from the ten triangle areas $j_f$ of $\s$ or when calculated from the ten triangle areas $j_f'$ of $\s'$, so that
\be L_\e^{(\s)}(j_1,\cdots, j_{10}) = L_\e^{(\s')}(j'_1,\cdots, j'_{10}) \,.\label{lc}\ee
This is precisely the constraint which turns an area-Regge action into a length-Regge action, see \cite{mak}.
Hence
\be S_{R}(j,\vec n)=\g\sum_f j_f (L) \d_f (L)= \frac{1}{8\pi l_p^2}\sum_f A_f (L)\d_f (L) \,,\label{lr}\ee
where the labels $L$ denote the lengths of the edges of the triangulation  and $l_P$ is the Planck length. 

If we refine infinitely the spacetime triangulation, then the Regge action in (\ref{lr}) will become the Einstein-Hilbert action $ S_{EH}/16 \pi l_P^2$. Therefore the ELPR/FK spin foam model with a modified amplitude (\ref{gva}) will give general relativity in the limit of large spins and smooth spacetime.

Note that the structure of the quantum corrections in the effective action (\ref{scea}) is the same as in the 3d case.

\bigskip
\bigskip
\noindent{\large\bf 6. Conclusions}

\bigskip
\noindent We have demonstrated that it is possible to construct spin foam models of quantum gravity in 3 and 4 spacetime dimensions such that the corresponding effective action has general relativity as its classical limit. The effective action method resolves the long-standing problem of how to compute the semiclassical limit of a spin foam model. The spin foam model defined by the vertex amplitude (\ref{gva}) is the first example of a finite four-dimensional spin foam model with the correct classical limit. 

In 3d case we have shown how to regularize the PR model and how to modify the vertex amplitude in order to obtain the correct classical limit. The same can be done in the case of the Turaev-Viro (TV) model \cite{tv}, which is a quantum group regularization of the PR model. The TV model effective action can be calculated since the large-spin asymptotics of the quantum $6j$-symbol is known \cite{mt}. Because the vertex asymptotics is also of the cosine type, this means that the TV model vertex amplitude has to be modified in order for the effective action to yield the EH action with a cosmological constant in the classical limit. The same applies to the quantum group regularization of the ELPR/FK model \cite{fm,h}, since the large spin asymptotics will be a deformation of the classical group asymptotics (\ref{eprlva}). We were also able to construct a Lorentzian version of the PR model such that the corresponding effective action has the desired classical limit, see (\ref{l3dqg}).

The structure of the quantum gravity corrections in 3 and 4 dimensions is the same; however, their spacetime interpretation is different. In 3d we have $j_f \propto L_\e$ so that the trace-log term in the 3d effective action is the discretization of the quantum field theory trace-log term, while the $\log j_f$ terms do not have the QFT analog. In 4d case, $j_f \propto A_f (L_\e)$ so that the hessian $S''_{R\,ff'}$ is not the same as the discretized version of the QFT hessian, which is $S''_{R\,\e\e'}$. Consequently the 4d effective action will have additional terms of the type $\log f(L_\e /l_P)$ where $f$ is a homogenous function of order one.
It would be interesting to analyze the implications of the $\log[ A_f (L_\e)/l_P^2 ]$ and $\log f(L_\e /l_P)$ terms for cosmology.

Note that the terms proportional to the $G_{ff'}$ matrix in (\ref{efac}) and (\ref{scea}) constitute  second-order quantum corrections, which are of $O(1/j)$. In order to obtain all second-order terms we need to know the $O(1/j)$ correction terms to the vertex Regge action in the vertex amplitude asymptotics.

It is not difficult to see that our formalism gives the area-Regge action as the classical limit of the appropriately modified Barret-Crane (BC) model, because the length constraint (\ref{lc}) is not enforced. Hence this justifies the concerns raised about the BC model, which were based on the convergence speed of the partition function, see \cite{ebcz}, as well as the concerns related to the graviton propagator for the BC model \cite{bcgp}.

Note that our method can be applied to any state sum model of quantum gravity, which means that one can study the effective action for the 4d Regge model.

\bigskip 
\bigskip
\noindent{\large\bf Acknowledgments}

\bigskip
\noindent
We would like to thank L. Freidel and J. W. Barrett for discussions. This work has been partially supported by FCT project PTDC/MAT/099880/2008. MV was also
supported by grant SFRH/BPD/46376/2008.

\end{document}